\documentstyle[prl,tighten,aps,epsf,graphics,multicol]{revtex}
\begin{document}
\draft
\title{Singlet-aided infinite resource reduction in the comparison
of distant fields}
\author{S. Bose$^{1}$, S.F.~Huelga$^{2}$ and M.B.~Plenio$^{3}$}
\address{${(1)}$ Center for Quantum Computation, University of Oxford,
Oxford OX1 3PU, United Kingdom,\\ ${(2)}$ Department of Physical
Sciences, University of Hertfordshire, Hatfield AL10 9AB, United
Kingdom,\\ ${(3)}$ Optics Section, The Blackett Laboratory,
Imperial College, London SW7 2BZ, United Kingdom.}
\date{\today}
\maketitle
\begin{abstract}
We present a task which can be faithfully solved with finite
resources only when aided by particles prepared in a particular
entangled state: the singlet state. The task consists of
identifying the mutual parallelity or orthogonality of weak
distant magnetic fields whose absolute directions are completely
unknown.
\end{abstract}
\begin{multicols}{2}
It is well known that quantum mechanics helps to reduce the
resources required to accomplish certain tasks
\cite{Deutsch,shor,grover1,Superdense,Cleve}. Some problems can be
solved with exponentially less resources when aided by quantum
mechanics, as featured in Shor's factorization algorithm
\cite{shor}. Other problems lead to a quadratic speed-up, such as
Grover's search algorithm \cite{grover1}. Moreover, the use of
quantum entanglement can result in resource reduction in a variety
of communication associated tasks. For example, in quantum dense
coding \cite{Superdense}, prior entanglement is used to increase
the classical information capacity of a quantum bit by a factor of
two. Sharing entanglement can also reduce the amount of
communication needed to evaluate certain functions of distributed
inputs \cite{Cleve}. In this paper, we present a task which
illustrates the superiority of an entanglement-based strategy in
very radical terms. We show that there is in fact an infinite gap
in the resources required for accomplishing the task with or
without the use of a certain entangled quantum state (a singlet).
Without sharing a singlet state, the task requires infinitely many
qubits for error free operation, while the use of shared singlets
reduces the resource requirement to at most four qubits.

    Consider the situation depicted in Figure 1., where two spatially
separated and disconnected regions are occupied by distant
partners Alice and Bob. A third person, Eve, who has access to
both separated zones, may subject these two regions to two weak
uniform magnetic fields of unit strength but otherwise random
direction. However, she gives Alice and Bob an important promise:
{\em the two fields are either parallel or orthogonal}. In other
words, if she had chosen the direction $\hat{n}$ for the field
applied on Alice's side, she chooses either $\hat{n}$ or any
direction $\hat{n}_{\scriptsize \perp}$ orthogonal to $\hat{n}$
for the field applied on Bob's side. We also make the assumption
that these fields are sufficiently weak so that they cannot be
determined classically.  The only way to determine the field
direction is by means of detecting their action on quantum states.
Alice  and Bob are given the task of {\em faultlessly identifying
(i.e. with unit probability of success), Eve's choice among the
two alternative relative orientations of the fields}. Note that
this is strictly a "quantum task" in contrast to existing examples
\cite{Deutsch,shor,grover1,Superdense,Cleve} in which quantum
mechanics is used to reduce the resources required to accomplish a
"classical task". Here, because of the weakness of the magnetic
fields, there is no hope to accomplish the task classically. But,
as we will show, even within the available quantum protocols,
using entanglement leads to an infinite resource reduction.

\begin{figure}
\begin{center}
\leavevmode \epsfxsize=7.cm
\epsfbox{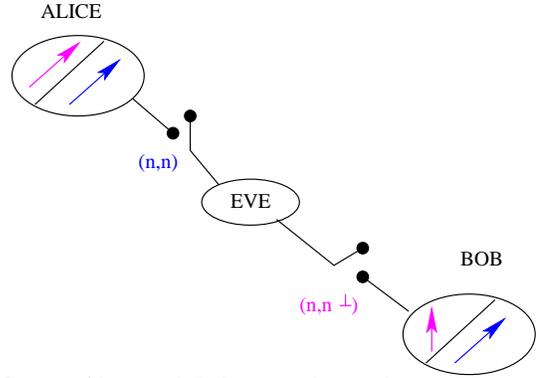}
\caption{\narrowtext \label{fields} Alice and Bob control two
distant regions. Eve may subject these two regions to a weak
magnetic field of unit strength but random direction. She gives
Alice and Bob only the promise that the two fields are either
parallel or orthogonal. Alice and Bob have the task of
distinguishing the two cases locally.}
\end{center}
\end{figure}

  We will first consider the case when Alice and Bob do not share
any entanglement. Suppose Alice has a qubit $A$ in an initial
state $|\psi\rangle_A$ and Bob has a qubit $B$ be in an initial
state $|\psi\rangle_B$. Let ${\vec
\sigma^{(A)}}=(\sigma^{(A)}_x,\sigma^{(A)}_y,\sigma^{(A)}_z)$ and
${\vec
\sigma^{(B)}}=(\sigma^{(B)}_x,\sigma^{(B)}_y,\sigma^{(B)}_z)$,
where $\sigma^{(A/B)}_i$ denotes the Pauli matrices of $A/B$. To
distinguish between the two alternatives locally would thus
require
\begin{equation}
\langle \psi| {\vec\sigma^{(B)}}.\hat{n} \otimes
{\vec\sigma^{(B)}}.\hat{n}_\perp |\psi\rangle_B =0.
\end{equation}
The only solution for this is for $|\psi\rangle_B$ to be an
eigenstate of the operator ${\vec\sigma^{(B)}}.\hat{n}$, where the
direction $\hat{n}$ is completely unknown. Therefore, in order to
account for error free detection, Bob will need to have a number
of qubits, each in an eigenstate of ${\vec\sigma^{(B)}}.\hat{n}$
corresponding to a different  $\hat{n}$. As there are an infinite
number of choices of $\hat{n}$, an error free detection scheme
requires Bob to hold an infinite number of qubits. It is important
to note that having any or all of Alice's or Bob's qubits in
classically correlated mixed states of the type $ \sum_i p_i
|\psi^i\rangle_A \langle \psi^i|_A \otimes |\psi^i\rangle_B
\langle \psi^i|_B$ will also not help in perfect discrimination of
${\vec\sigma^{(A)}}.\hat{n}\otimes {\vec\sigma^{(B)}}.\hat{n}$ and
${\vec\sigma^{(A)}}.\hat{n}\otimes
{\vec\sigma^{(B)}}.\hat{n}_\perp$. If that were the case, then one
would have been able to choose three mutually perpendicular
directions and perform quantum dense coding of capacity
$\log_2{3}$ bits per qubit. But this is not possible with a
disentangled state, as shown in Ref.\cite{mdc}.

    Let us now describe a strategy where Alice and Bob initially share
entanglement. Imagine that the qubits $A$ and $B$ possessed by
Alice and Bob are prepared in a singlet state
\begin{equation}
    |\psi^{-}\rangle = \frac{1}{\sqrt{2}} ( |01\rangle - |10\rangle ) .
    \label{Eq1}
\end{equation}
Now Eve subjects Alice's and Bob's qubits to her chosen unitary
transformations. Suppose she chose the pair $\{ {\vec
\sigma^{(A)}}.\hat{n},{\vec \sigma^{(B)}}.\hat{n} \}$, (i.e.
parallel fields). Then the state shared by Alice and Bob evolves
to
\begin{eqnarray}
    |\psi\rangle &=& {\vec \sigma^{(A)}}.\hat{n} \otimes {\vec \sigma^{(B)}}.\hat{n}
    |\psi^-\rangle \nonumber \\
   &=& |\psi^-\rangle,
    \label{Eq4}
\end{eqnarray}
where we have used the fact that a singlet state is invariant
under operations $U^{(A)}\otimes U^{(B)}$ (i.e. when the same
unitary operation $U$ is applied to both qubits). On the other
hand, if Eve decided to apply the pair $\{ {\vec
\sigma^{(A)}}.\hat{n},{\vec \sigma^{(B)}}.{\hat{n}_\perp} \}$
(i.e. perpendicular fields) the singlet will evolve to a coherent
superposition of the three triplet states $|\psi^{+}\rangle,
|\Phi^{+}\rangle$, and $|\Phi^{-}\rangle$. This can easily be seen
from the fact there is always a unitary transformation
$U(\hat{n})$ such that
\begin{equation}\label{Eq6}
    {\vec \sigma^{(A)}}.\hat{n} \otimes {\vec \sigma^{(B)}}.\hat{n}_\perp = U(\hat{n}) \sigma^{(A)}_x U(\hat{n})^{\dagger}
    \otimes U(\hat{n}) \sigma^{(B)}_{y/z} U(\hat{n})^{\dagger}
\end{equation}
and therefore
\begin{equation}\label{Eq7}
  \langle \psi^{-}| {\vec \sigma^{(A)}}.\hat{n} \otimes {\vec \sigma^{(B)}}.\hat{n}_\perp
  |\psi^{-}\rangle = \langle \psi^{-}| \sigma^{(A)}_x \otimes \sigma^{(B)}_{y/z}
  |\psi^{-}\rangle = 0.
\end{equation}
As a result, Alice and Bob can now easily check which of the two
possible relative orientations Eve has chosen. If the parallel
configuration $\{ {\vec \sigma^{(A)}}.\hat{n},{\vec
\sigma^{(B)}}.\hat{n} \}$ was applied, Alice and Bob still share a
singlet state. On the contrary, if they were subject to the
orthogonal configuration $\{ {\vec \sigma^{(A)}}.\hat{n},{\vec
\sigma^{(B)}}.\hat{n}_\perp \}$, Alice and Bob now hold a state
that is orthogonal to the singlet state.

\begin{figure}
\begin{center}
\leavevmode \epsfxsize=7cm
\epsfbox{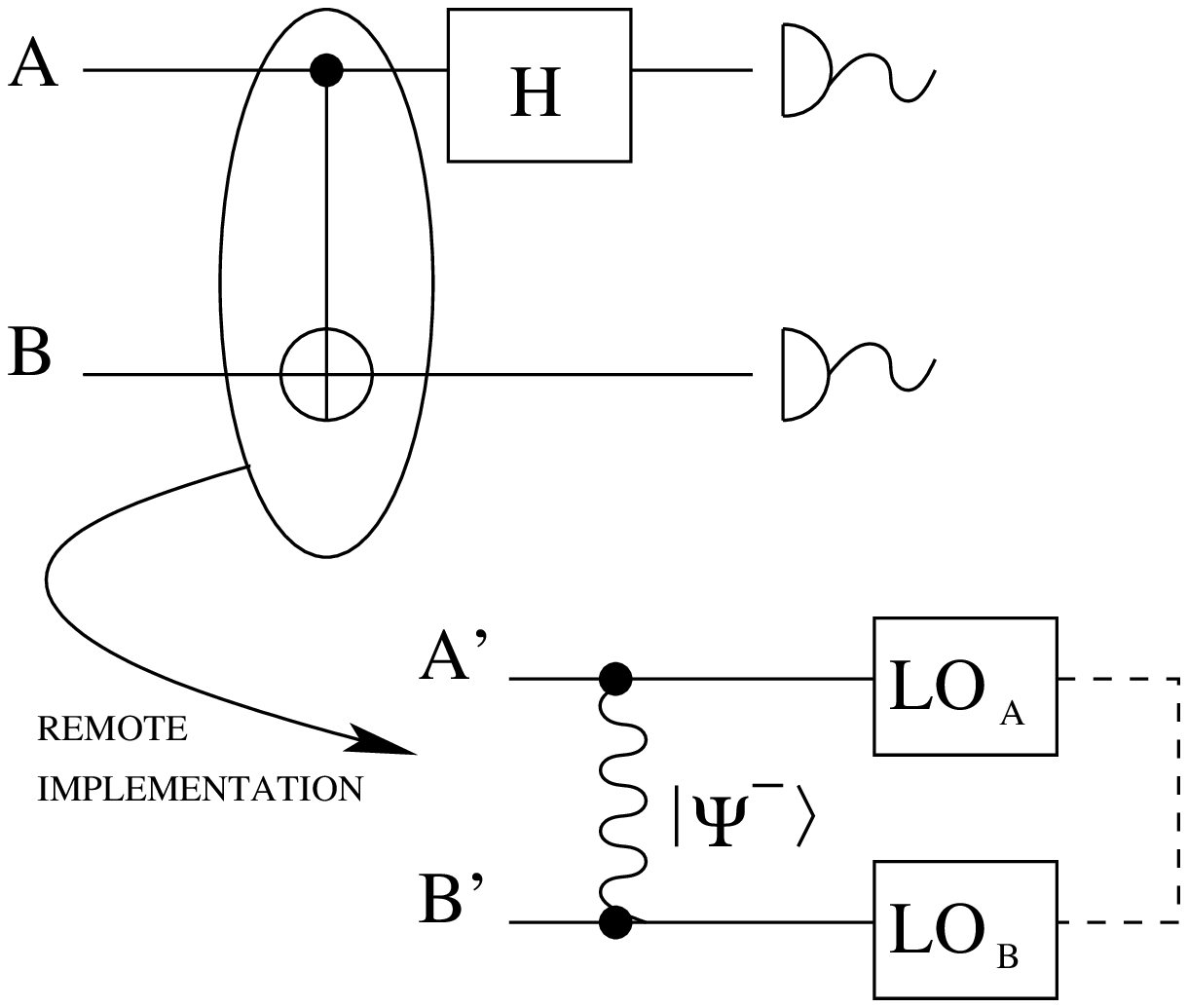}
\label{measurement} \caption{\narrowtext  A controlled-NOT and an
inverse Hadamard suffice to determine whether one has a singlet
state or a triplet state. A controlled-NOT in turn can be
implemented consuming one extra singlet \protect\cite{eisert}
which has been shielded from the action of the magnetic field and
using only local operations (LO) supplemented by classical
communication, as illustrated by the dashed line in the diagram.}
\end{center}
\end{figure}

  For the determination which state Alice and Bob are holding, two
scenaria are possible. In one, we may assume that Alice and Bob
are allowed to send each other quantum particles. In this case
Alice simply sends Bob her particle, and Bob then measures the
projection operator onto the singlet space. If this projection is
successful, then he knows that the fields were parallel, if the
projection was not successful then the fields were orthogonal.
However, one may also demand that Alice and Bob only share some
initial entanglement in the form of singlets and that no further
quantum communication is possible. In that case Alice and Bob need
altogether two pairs of singlet states to complete the task. The
first pair is treated as outlined above, while the second one is
kept isolated from Eve and will be needed to determine whether the
first pair is in a singlet state or not. This can clearly be done,
as one singlet state is enough to implement a controlled-NOT gate
remotely \cite{eisert}. The quantum circuit required for the
(local) discrimination of the shared entangled state is shown in
Figure 2. First a remote controlled-NOT gate with Alice's qubit
acting the control bit is applied. This process consumes an e-bit
of entanglement. Subsequently a Hadamard transformation onto
Alices qubit takes the state $|0\rangle$ into $|0\rangle -
|1\rangle$ and the state $|1\rangle$ into $|0\rangle + |1\rangle$.
As a result of this protocol, the four Bell states are mapped into
product states as follows
\begin{eqnarray}
    |00\rangle + |11\rangle &\longrightarrow& |00\rangle \\
    |00\rangle - |11\rangle &\longrightarrow& |10\rangle \\
    |01\rangle + |10\rangle &\longrightarrow& |01\rangle \\
    |01\rangle - |10\rangle &\longrightarrow& |11\rangle .
\end{eqnarray}
The state discrimination is completed when Alice and Bob measure
their first pair. If they both find the state $|1\rangle$, then
they initially shared a singlet state and the magnetic fields were
parallel. Therefore, even without the use of quantum
communication, Alice and Bob require only two singlets to
accomplish the task of determining the relative orientation of the
two fields without error.

     We will now prove that the singlet is the {\em only} state which
allows Alice and Bob to achieve error free discrimination. The
state $|\psi\rangle$ which Alice and Bob must share in order to
accomplish the required task  needs to satisfy
\begin{equation}
\label{singl} \langle \psi| {\vec\sigma^{(A)}}.\hat{n} \otimes
{\vec\sigma^{(B)}}.\hat{n}~{\vec\sigma^{(A)}}.\hat{n}^{'} \otimes
{\vec\sigma^{(B)}}.\hat{n}^{'}_\perp |\psi \rangle =0,
\end{equation}
where $\hat{n}$ and $\hat{n}^{'}$ are two {\em completely
arbitrary} directions. This is because Alice and Bob need to
distinguish {\em all cases} of parallel field directions from {\em
all cases} of perpendicular field directions.  In other words, Eve
could have picked one absolute direction $\hat{n}$ to apply
parallel magnetic fields and a different absolute direction
$\hat{n}^{'}$  to apply the perpendicular fields. These two cases
need to be perfectly distinguished. We first find out the
restrictions on $|\psi\rangle$ which already arise from
considering the special case $\hat{n}=\hat{n}^{'}$. In that case
Eq.(\ref{singl}) simplifies to
\begin{equation}
\label{eql} \langle \psi| I^{A} \otimes
{\vec\sigma^{(B)}}.\hat{n}^{''} |\psi \rangle =0,
\end{equation}
where $I^{A}$ is the identity operator for qubit $A$ and
$\hat{n}^{''}=\hat{n}\times\hat{n}_\perp$ (i.e. $\hat{n}^{''}$ is
arbitrary as $\hat{n}$ itself is arbitrary). Eq.(\ref{eql})
restricts the class of allowed $|\psi\rangle$ to maximally
entangled states of the qubits $A$ and $B$. If we put a maximally
entangled state $|\psi\rangle_{\mbox{max}}$ in Eq.(\ref{singl})
and simplify, we get the condition
\begin{eqnarray}
(\hat{n}.\hat{n}^{'})(\hat{n}.\hat{n}^{'}_\perp)&-&\langle\psi|{\vec\sigma^{(A)}}.(\hat{n}\times
\hat{n}^{'})\nonumber \\\otimes
{\vec\sigma^{(B)}}.(\hat{n}&\times& \hat{n}^{'}_\perp)|\psi\rangle
_{\mbox{max}}=0.
\end{eqnarray}
Substituting $|\psi\rangle_{\mbox{max}}$ in the above equation by
its expansion
$c_1|\psi^{+}\rangle+c_2|\psi^{-}\rangle+c_3|\phi^{+}\rangle
+c_4|\phi^{-}\rangle$ in terms of Bell states gives the unique
solution $|c_2|^2=1$. This proves that the only state which
satisfies Eq.(\ref{singl}) is the singlet state. Thus {\em the
singlet becomes the only feasible state for error free
discrimination}.

   Let us now briefly point out how our scheme differs from those schemes
which appear to be related. The fact that our scheme can only be
carried out with singlets makes it different from quantum dense
coding \cite{Superdense}
 and precision magnetic field determination \cite{Preskill},
which would both work for {\em any} maximally entangled state. It
is also different from the standard inability to distinguish
specific entangled (and also some unentangled \cite{nwe}) states
locally by finite resources. Here, we are not given any prior sets
of unknown states to distinguish, but some unknown relative
orientations of fields to discriminate. We have identified the
state which works best for this purpose (namely the singlet).

   The task
of determining the relative orientation of two magnetic fields can
be generalized in many directions. First of all one may allow for
more possible relative directions, and ask Alice and Bob to
determine the angle between the two directions. In this case both
the entanglement based as well as the disentangled strategy are
unable to deliver error free answers, but it can be expected that
the entangled strategy will deliver the better overall precision
or the lower error rate. A further generalization of the problem
could also allow for a variable strength of the magnetic field.
Again it can be expected that the entanglement based strategy will
be superior. It should be noted that this problem is related to
that of atomic frequency standards \cite{Frequencystandards},
which can be mapped onto a problem where a field of known
orientation but unknown strength should be detected with the best
possible resolution. It should also be noted that Eve could have
given a different promise leading to a similar gap between the
entanglement based and the disentangled strategy. If Eve promises
that the fields are either parallel or anti-parallel, then either
the singlet state remains invariant, or it is converted (after a
suitable waiting time) into the triplet state $|01\rangle +
|10\rangle$, which in turn allows to determine the relative
orientation of the fields. Summarizing, we have presented a task
that be solved efficiently using shared entanglement in the form
of singlet states and demonstrated that the associated cost in
resources represents an infinite gap as compared to applying a
classical strategy. Error free performance requires that Alice and
Bob hold an infinite number of disentangled particles while an
entanglement-based strategy uses either one singlet pair, if
subsequent quantum communication is allowed, or, at most, two
singlet pairs if only local quantum operations and the exchange of
classical communication is allowed. It is quite interesting to
note that very recently yet another application of entanglement
which uses the $U\otimes U$ invariance of a singlet in an
essential way has been proposed \cite{jos}.

   This work has been supported by The Leverhulme Trust,
the EQUIP project of the European Union, the UK Engineering and
Physical Sciences Research Council (EPSRC) and DGICYT Project
No.~PB-98-0191 (Spain).

\end{multicols}

\begin{references}
%
\bibitem{Deutsch} D. Deutsch, Proc. Roy. Soc. A {400}, 97 (1985);
D. Deutsch and R. Jozsa, Proc. Roy. Soc. Lond. A {\bf 439}, 553
(1992).
%
\bibitem{shor} P.W.~Shor, in {\em Proceedings of the 35th Annual
Symposium on the Theory of Computer Science\/}, edited by S.
Goldwasser (IEEE Computer Society Press, Los Alamitos, CA), p.124
(1994).
%
\bibitem{grover1} L.K.~ Grover, Phys. Rev. Lett. {\bf 79}, 325 (1997).
%
%
\bibitem{Superdense} C.H. Bennett and S.J. Wiesner, Phys. Rev. Lett. {\bf 69},
2881 (1992).
%
\bibitem{Cleve} R. Cleve and H. Buhrman, Phys. Rev. A {\bf 56}, 1201
(1997).
%
%
\bibitem{mdc} S. Bose, M.B. Plenio and V. Vedral, J. Mod. Optics {\bf 47}, 291 (2000).
%
\bibitem{eisert} J. Eisert, K. Jacobs, P. Papadopoulos and M.B.~Plenio
(to be published).
%
\bibitem{Preskill} A. Childs, J. Preskill, and J. Renes, J. Mod. Optics {\bf 47},
155 (2000)
%
\bibitem{nwe}
C.H. Bennett, D.P. DiVincenzo, C.A. Fuchs, T. Mor, E. Rains, P.W.
Shor, J.A. Smolin and W. K. Wootters, Phys. Rev. A {\bf 59}, 1070
(1999).
%
\bibitem{Frequencystandards}W. J. Wineland et al., {Phys. Rev.} A{\bf 46}, R6797
(1992); S.F.~Huelga, C. Macchiavello, T. Pellizzari, A.K.~ Ekert,
M.B. Plenio and J.I. Cirac, Phys. Rev. Lett. {\bf 79}, 3865
(1997).
%
\bibitem{jos}
R. Jozsa, D. S. Abrams, J. P. Dowling and C. P. Williams, {\em
Quantum atomic clock synchronization based on shared prior
entanglement}, LANL eprint quant-ph/0004105.
\end{references}
\end{document}